\documentclass[12pt]{iopart}
\usepackage{graphicx}

\usepackage{iopams} 
\usepackage{color}
 
\begin{document}

\title{Crossover of skyrmion and helical modulations \\
in noncentrosymmetric ferromagnets
}

\author{Andrey O. Leonov and Alexei N.\ Bogdanov }

\address{Department of Chemistry, Faculty of Science, Hiroshima University, Kagamiyma, 
Higashi Hiroshima 739-8526, Japan}

\address{Chirality Research Center, Hiroshima University, Higashi Hiroshima, 
Hiroshima 739-8526, Japan}

\address{IFW Dresden, Postfach 270116, D-01171 Dresden, Germany}

\ead{leonov@hiroshima-u.ac.jp}

\ead{a.bogdanov@ifw-dresden.de}

\begin{abstract}
{The coupling between angular (twisting) and longitudinal 
modulations arising near the ordering temperature of noncentrosymmetric 
ferromagnets strongly influences the structure of skyrmion states and their 
evolution in an applied magnetic field. 
In the precursor states of cubic helimagnets, a continuous transformation 
of  skyrmion lattices into the saturated state is replaced by 
the first-order processes accompanied by the formation of multidomain
states.
Recently the effects imposed by dominant longitudinal modulations 
have been reported in bulk MnSi and FeGe. Similar phenomena can be observed 
in the precursor regions of cubic helimagnet epilayers and in easy-plane chiral 
ferromagnets (e.g. in the hexagonal helimagnet CrNb$_3$S$_6$).
}
\end{abstract}

         
 \maketitle

\begin{figure}
\includegraphics[width=1.0\columnwidth]{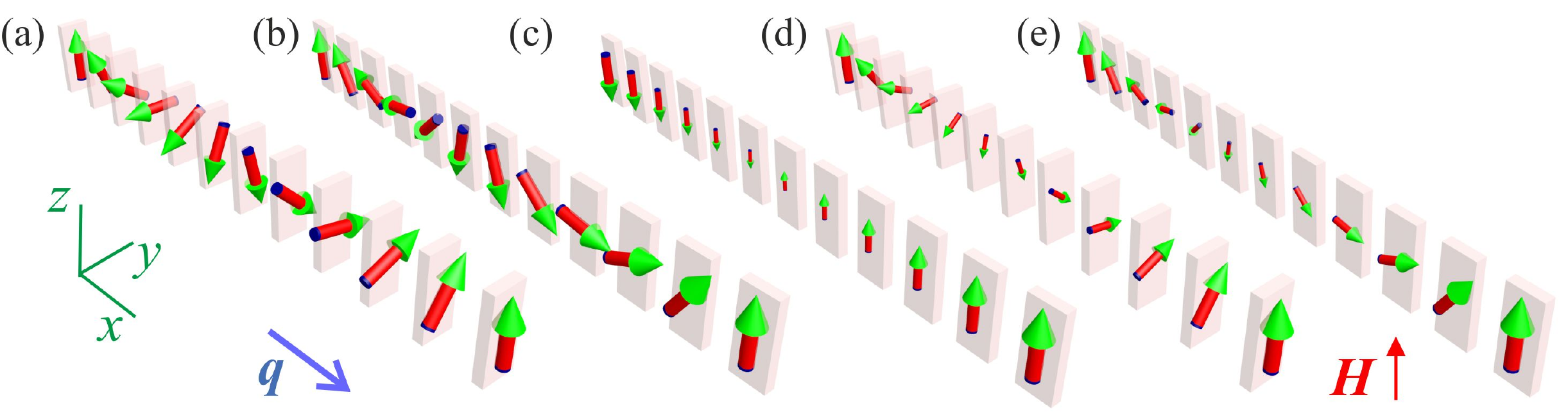}
\caption{ 
(color online) Arrays with different types of modulations: chiral modulations with 
the fixed magnetization magnitude $M$ of Bloch (a) and N$\acute{e}$el type (b), longitudinal 
modulations of $M$ (c), combined longitudinal  and angular modulations of Bloch (d) 
and N$\acute{e}$el (e) helices in the applied magnetic field along the \textit{z}-axis.
\label{fig:arrays}
}
\end{figure}

\textit{Introduction}. In magnetically ordered noncentrosymmetric 
crystals the chiral asymmetry of the exchange interaction induces specific 
\textit{Dzyaloshinskii-Moriya} (DM) interactions \cite{Dz64}.
In the energy functional of noncentrosymmetric ferromagnets, 
these interactions are described by energy contribution with antisymmetric 
terms linear in the first spatial derivatives of the magnetization
$\mathbf{M} (\mathbf{r})$ (\textit{Lifshitz} invariants) \cite{Dz64}

\begin{eqnarray}
 \mathcal{L}_{ij}^{(k)} = M_i \frac{\partial M_j}{\partial x_k} 
- M_j \frac{\partial M_i}{\partial x_k}  
\label{lifshitz}
\end{eqnarray}
where $x_k$ are Cortesian components of the spatial variable $\mathbf{r}$.
Energy functionals $\mathcal{L}_{ij}^{(k)}$ (\ref{lifshitz}) 
favour  spatial modulations of the magnetization propagating along 
the $x_k$ axis and  rotating in the ($x_i, x_j$) plane with a \textit{fixed} 
rotation sense. 
 For instance,  the energy functionals $\mathcal{L}_{yz}^{(x)}$ 
and $\mathcal{L}_{xz}^{(x)}$ stabilize Bloch and  N$\acute{e}$el type modulations 
correspondingly (figure \ref{fig:arrays} (a), (b)). 
The competition of the DM interaction with other magnetic forces 
stabilizes spatially modulated phases in a form of one-dimensional 
\textit{helices} \cite{Dz64} or two-dimensional
modulations composed of axisymmetric cells called \textit{skyrmion lattices}
\cite{JETP89,JMMM94}. The magnetic-field-driven evolution of helical 
and skyrmionic states has been discovered and investigated in several 
groups of magnetic compounds with intrinsic and induced chirality.
 The former include noncentrosymmetric magnetically ordered crystals
(e.g. cubic and uniaxial helimagnets \cite{Yu10,Togawa12PRL})
and the latter nanolayers of achiral ferromagnetic metals with surface/interface 
induced DM interactions (e.g. FePd/Ir(111) nanolayers \cite{Leonov16NJP}).

In ferromagnetic materials at a fixed temperature, the magnetization modulus
$M = |\mathbf{M}|$ remains constant and is independent of the value of
 applied magnetic field.
However, near the ordering temperature the longitudinal stiffness decreases
and spatial longitudinal modulations of the magnetization modulus 
become a sizable effect (Fig. \ref{fig:arrays} (c)).
As a manifestation of this effect we recall the instability of rotational
(Bloch and N$\acute{e}$el) domains  walls  near the ordering temperatures of uniaxial
ferromagnets, and the formation of domain walls with longitudinal modulations
of the magnetization amplitude \cite{Ginzburg64,Koetzler93}.

For magnetic  skyrmions the ``softening'' of the magnetization modulus
has dramatic consequences.  Near the ordering temperature in applied magnetic 
fields, skyrmionic textures consist of complex combinations of rotational 
and longitudinal modulations (figure \ref{fig:arrays} (d), (e))
\cite{Leonov10condmat,Wilhelm11,Wilhelm12}.
It was also established that the coupling of rotational 
and longitudinal modes strongly influences the structure and magnetic 
properties of skyrmionic states and  leads to the \textit{crossover} of 
the interparticle \textit{skyrmion-skyrmion} interactions and a specific
confinement effects of skyrmion lattices \cite{Leonov10condmat, Wilhelm11,Wilhelm12}.
Earlier similar effects have been investigated in one-dimensional
chiral modulations (\textit{helicoids}) \cite{Mukamel85,Yamashita87}.
In \cite{Wilhelm11,Wilhelm12} we have introduced a characteristic \textit{crossover
temperature } $T_{\mathrm{cf}}$ ($(T_{\mathrm{C}} - T_{\mathrm{cf}})/T_{\mathrm{C}} \ll 1$)
separating the temperature interval below the Curie temperature ( $T < T_{\mathrm{C}}$)
into two distinct regions. Over the broad temperature  range from 0 to $T_{\mathrm{cf}}$
chiral modulated textures are composed of  rotational modes (so called \textit{regular}
modulations). In the narrow temperature interval ($T_{\mathrm{cf}} < T < T_{C}$),
specific \textit{precursor} modulations with a strong coupling between
rotational and longitudinal modes determine the peculiar magnetic properties of chiral
ferromagnets in this region.

During the last years, the precursor states of noncentrosymmetric ferromagnets 
have been investigated experimentally and theoretically 
\cite{Laliena16,Shinozaki17,Moskvin13prl,Wilhelm16,Pappas17prlmnsi,XuAPL17FeGe,
Tsuruta16PRB,ClementsCrNbS,Turgut17,Siegfried17}.
Particularly, the recent observations in bulk cubic helimagnets 
MnSi \cite{Pappas17prlmnsi} and FeGe \cite{Moskvin13prl,Wilhelm16,Pappas17prlmnsi} 
are consistent with the crossover effects of skyrmionic states theoretically described
in \cite{Leonov10condmat, Wilhelm11,Wilhelm12} and earlier reported
in a cubic helimagnet FeGe \cite{Wilhelm11,Wilhelm12}  

In this paper we develop a consistent theory of skyrmion states
near the ordering temperature and construct a magnetic phase 
diagram based on numerical calculations of the equilibrium skyrmionic 
states describing  the crossover region between
regular and precursor modulations. We discuss the results of
 \cite{Moskvin13prl,XuAPL17FeGe,Pappas17prlmnsi} and the peculiarities
of the precursor skyrmions in nanolayers of cubic and uniaxial
chiral ferromagnets.

 \textit{Regular and precursor skyrmion states}.  
For investigations of the crossover phenomenon we consider the stardard isotropic model 
for cubic noncentrosymmetric ferromagnets near the ordering temperature 
\cite{Bak80,Wilhelm11} 
\begin{equation}
W= A \left(  \mathrm{grad} \: \mathbf{M}\right)^2+D \mathbf{M} \cdot \mathrm{rot} \mathbf{M} 
-\mathbf{M}\cdot\mathbf{H}+ W_0 (M)
\label{density}
\end{equation}
where  $A$ is the exchange stiffness, $D$ is the Dzyaloshinskii
constant, $\mathbf{H}$ is an applied magnetic
field, $W_0$ collects short-range magnetic interactions dependent
on the magnetization modulus $M = |\mathbf{M}|$. Near the ordering
temperature  $W_0$ can be written as \cite{Bak80} 
\begin{eqnarray}
W_0 =  \alpha (T-T_0) M^2 + b M^4, \quad b > 0
\label{densitya}
\end{eqnarray}
where $T_0$ is the ordering temperature of
a ferromagnet with $D = 0$.

The functional (\ref{density}) contains only the basic interactions 
 essential to stabilize skyrmionic states in noncentrosymmetric
ferromagnets. In particular, this includes the DM energy in the most symmetric 
\textit{isotropic} form, and, thus, describes the most general properties of
chiral modulations common for all noncentrosymmetric ferromagnets.
 In this paper, we omit some less important energy contributions 
such as magnetic anisotropy, stray-field energy, magneto-elastic coupling.

At zero field, model (\ref{density}) describes the formation of one-dimensional
twisting modulations (\textit{helices}) (figure \ref{fig:arrays} (a))
 with period $L_D$ below the Curie temperature, 
$T_{\mathrm{C}}$ where \cite{Bak80,Wilhelm12}  
\begin{eqnarray}
L_D = \frac{4 \pi A} {|D|}, \quad  T_C = T_0 +\Delta_D/4, \quad  T_{cf} = T_C -\Delta_D,
\quad  \Delta_D = \frac{D^2}{2\alpha A}. \quad
\label{units0}
\end{eqnarray}
\textit{Exchange shift} $\Delta_D$ (\ref{units0})
is a characteristic temperature related to 
the difference between the Curie temperature of a chiral
ferromagnet($T_C$) and $T_0$. 
We introduce here the \textit{crossover} temperature
$T_{cf}$ separating the temperature intervals with \textit{repulsive} ($0 < T < T_{cf}$) and 
\textit{attractive} ($T_{cf} < T < T_C$) skyrmion-skyrmion interactions 
\cite{Leonov10condmat,Wilhelm11,Wilhelm12}
The connection with the inter-skyrmion coupling crossover and the properties 
of skyrmion lattices will be discussed at the end of the paper.

By rescaling the spatial variable $\mathbf{r} = \mathbf{x}/r_0$ 
where $r_0 = L_D/(4 \pi)$ is related
to the helix period $L_D$ (\ref{units0}), the energy density functional 
(\ref{density}) is reduced to the following form 
\begin{eqnarray}
\Phi_1  =(\mathbf{grad}\: \mathbf{m})^2
- \mathbf{m}\cdot \mathrm{rot} \mathbf{m}
-h(\mathbf{k} \cdot \mathbf{m}) + (t+1/4) m^2+ m^4.
\label{density2}
\end{eqnarray}
We introduce here the reduced temperature,
magnetic field, and magnetization
\begin{eqnarray}
t = (T-T_C)/\Delta_D , \quad  \mathbf{h} = \mathbf{H}/H_0, \quad
 \mathbf{m} =  \mathbf{M}/M_0, \quad
\label{units1}
\end{eqnarray}
where
$H_0 = \alpha \Delta_D M_0$, $M_0 = \sqrt{\alpha \Delta_D/b }$,
 $h = |\mathbf{h}|$, and $\mathbf{k}$ is a unity vector along 
the applied magnetic field.

\begin{figure}
\includegraphics[width=0.95\columnwidth]{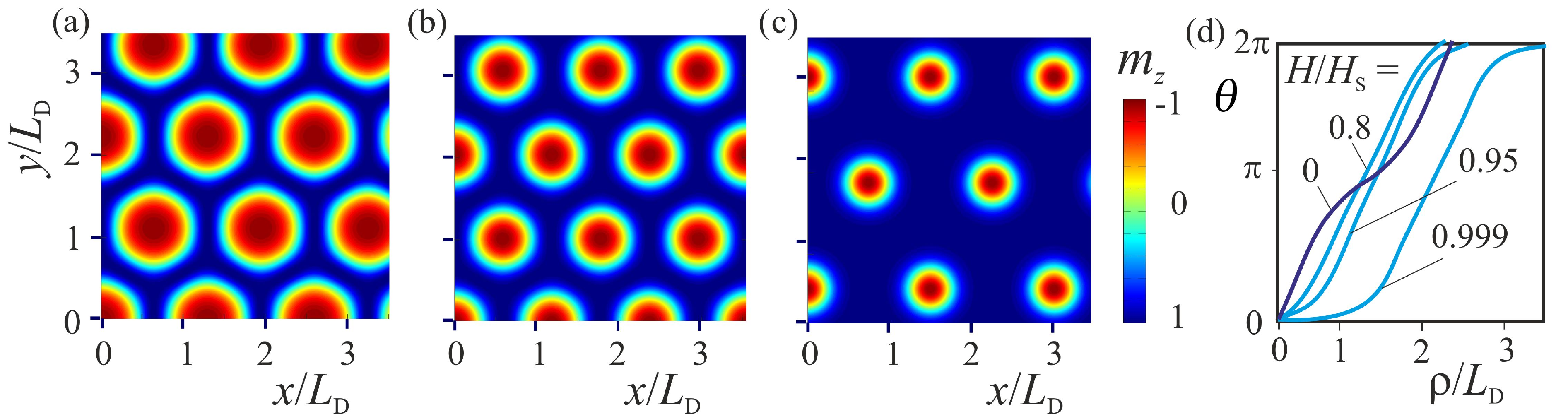}
\caption{ 
(color online)   { A hexagonal skyrmion lattice 
with a fixed magnetization magnitude below the critical field
$H_s = 0.801 H_D$ \cite{JMMM94}.
The calculated contour plots of  $m_z (x,y)$ for $H/H_s$ = 0 (a), 0.5 (b), 0.999 (c),
and magnetization profiles $\theta (\rho)$ (d) describe a magnetic-field-driven evolution
of a skyrmion lattice. At $H = H_s$ the skyrmion lattice
transforms into a set of repulsive isolated skyrmions \cite{JMMM94}.} 
\label{fig:unwinding}
}
\end{figure}

\begin{figure}
\includegraphics[width=0.8\columnwidth]{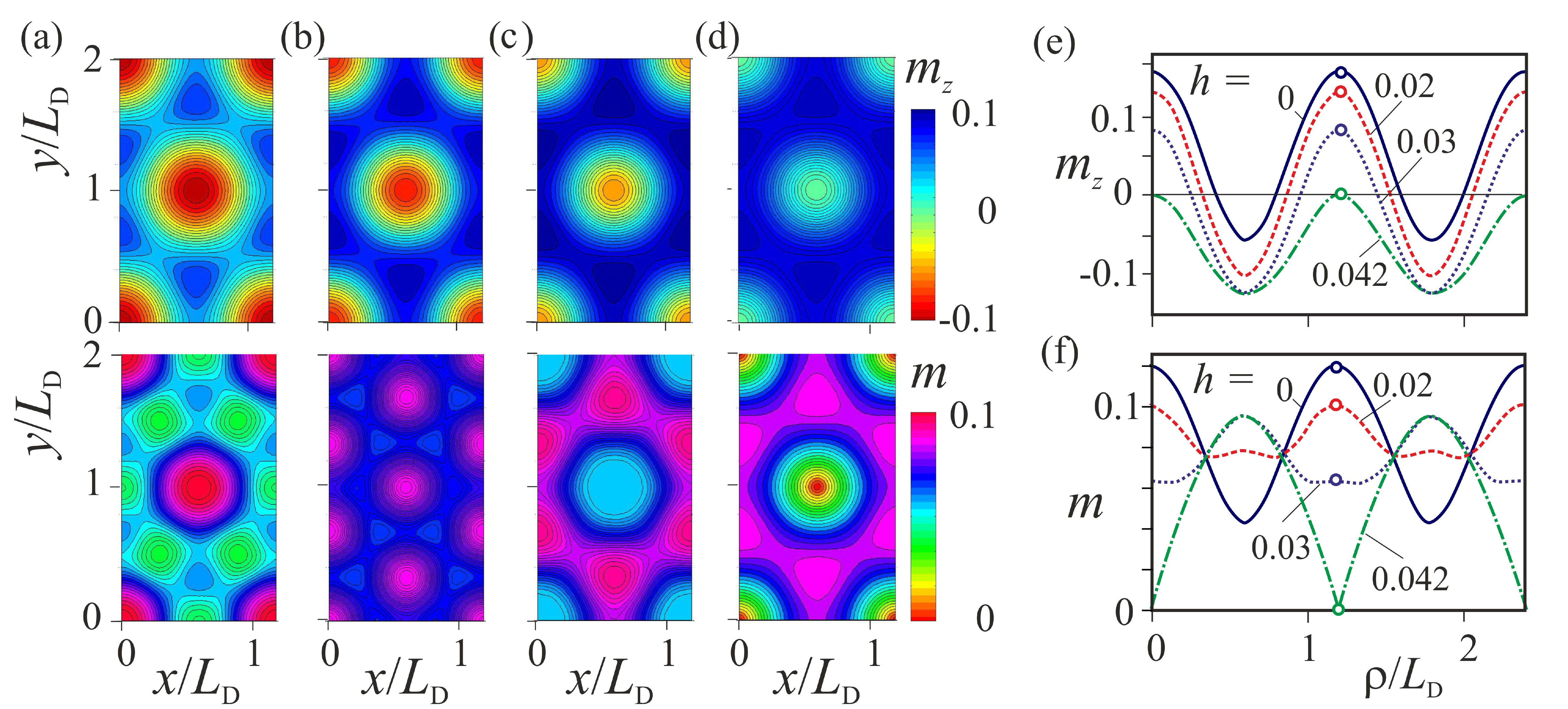}
\caption{ 
(color online) Contour plots $m_z (x,y)$ (top) 
and $m (x,y)$ (bottom) for 
a skyrmion lattice near the ordering temperature ($t = -0.02$) 
and different values of the applied magnetic field 
( $h = $  0 (a), 0.02 (b), 0.03 (c), 0.042 (d)).
 Diagonal cross-sections $m_z (\rho)$ (e) 
and $m (\rho)$ (f) for the skyrmion lattice core at $t = -0.02$.
\label{fig:skyrmion2}
}
\end{figure}

Spatial variations of the magnetization modulus occurs
only in  the vicinity of the ordering temperature.
Beyond this region at  fixed temperatures,  the modulus is equal 
to the saturated value, $M = M_s (T)$ within the whole sample 
and independent of the applied magnetic field.
For this case, the functional (\ref{density2}) can be replaced by
the following reduced functional
\begin{eqnarray}
\Phi_2  =(\mathbf{grad}\: \mathbf{n})^2
- \mathbf{n}\cdot \mathrm{rot} \mathbf{n}
-(H/H_D) (\mathbf{k} \cdot \mathbf{n}), \ 
\ \
\label{density1}
\end{eqnarray}
where $H_D (T)  = D^2 M_s (T)/(2A)$ is the saturation field \cite{Wilhelm12,Butenko10},
  $\mathbf{n}$ and $\mathbf{k}= \mathbf{H}/|\mathbf{H}|$ are  unity vectors 
along the magnetization $\mathbf{M}$ the applied magnetic field $\mathbf{H}$ 
correspondingly. 
Note that  models (\ref{density2}) and (\ref{density1}) have different scales 
for  applied magnetic fields. Rotational modulations within 
the model with a fixed modulus (\ref{density1}) are described by the reduced field $H/H_D$, 
and the equations for magnetic modulations with a ``soft'' modulus inlclude the reduced magnetic
field $H/H_0$ (\ref{density2}).

We consider solutions for hexagonal skyrmion lattices which are homogeneous along the skyrmion
axis ($z$ axis in this paper). This assumption is violated near the surfaces of cubic
helimagnets where twisting modulations with the propagation along the $z$-axis
arise \cite{Meynell14prb}. However, these modulations are strongly localized 
near the surfaces (their penetration depth is estimated as $0.1 L_D$ \cite{Meynell14prb}) 
and are ignored in our work.

For the model with a soft magnetization modulus the equilibrium parameters 
for hexagonal skyrmion cells $\mathbf{m} (x,y) $ are derived by minimization 
of functional $\Phi_1$ (\ref{density2}). 
This depends on three internal variables (components of the magnetization vector
$\mathbf{m}$) and two control parameters, the \textit{reduced magnetic field amplitude}
$h$ and the  \textit{"effective" temperature $t$}.

The solutions for the model with a fixed modulus $\mathbf{n} (x,y)$ 
i.e. well below the ordering temperature are derived
by minimization of functional $\Phi_2$ (\ref{density1}).
Numerical solutions are obtained by using finite differences 
for gradient terms and adjustable grids to accommmodate modulated 
states with periodic boundary conditions \cite{Wilhelm11}.

An important insight into the properties of skyrmion lattices is gained by 
calculations within the circular cell approximation \cite{JMMM94,Nature06}. 
In this method  the hexagonal cell is replaced by a circle of
equal area transferring the boundary conditions at the cell
edge from the hexagon to the circle.    
Following \cite{JMMM94,Nature06} we use spherical coordinates for
$\mathbf{m} (x,y)$ and cylindrical for the spatial variable ($\mathbf{x}$):

$$\mathbf{m} = m(\sin \theta \cos \psi, \sin \theta \sin \psi, \cos \theta), \quad 
\mathbf{x} = (\rho \cos \varphi, \rho \sin \varphi, z)$$

and consider a "lattice" composed of circular cells
of radius $R$.
After substitution of solutions $\psi = \varphi + \pi/2$
into (\ref{density2}) and integration with respect to
$\varphi$ the energy density  per unit volume $\mathcal{W}(R)$ for
the lattice in the circular cell approximation is
$\mathcal{W}(R) = (2/R^2) \int_{0}^{R} w (\theta, m, \rho) \rho d \rho$. 
The Euler equations for the functional $\mathcal{W}(R)$
with boundary conditions $\theta(0) = \pi$, $\theta(R) = 0$,
$m(R) = m_R $ describe the solutions for a Skyrmion cell
and yield the energy density $\mathcal{W}$ as a 
function of the control parameters $t$, $h$ and
the internal parameters $m_R$ and $R$. 
The equilibrium profiles $\theta (t, h)$, $m (t, h)$
are derived by the minimization of $\mathcal{W} (m_R, R)$
with respect to $m_R$, $R$.
A similar procedure is applied for calculations within 
the model with a fixed magnetization modulus \cite{JMMM94}.

\begin{figure}
\includegraphics[width=0.8 \columnwidth]{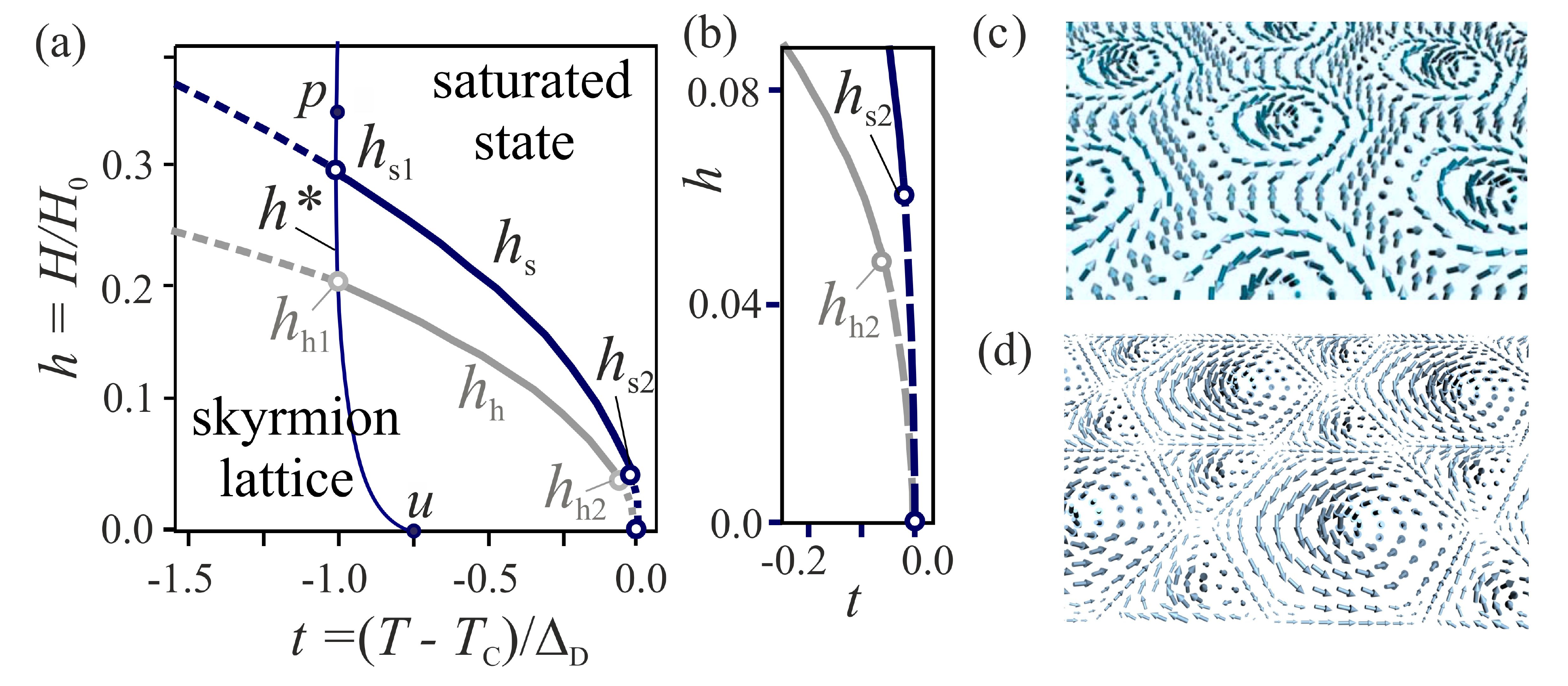}
\caption{
(color online).
(a) The phase diagram  in reduced variables of temperature
$t = (T - T_C)/\Delta_D$ and applied magnetic field $h = H/H_0$ (Eq. (6))
indicate the transition lines $h_s(t)$ from the skyrmion lattice into 
the saturated state (a). $h* $ is the critical line (8).
 For $h_s >h_{s1}$ the repulsive skyrmion cores
transforms into a set of isolated skyrmions by the second-order transition
of nucleation type (figure \ref{fig:unwinding}).
The first-order transition occurs for $h_{s1} < h_s < h_{s2}$ 
(figure \ref{fig:skyrmion2}). 
For $ h_{s2} > h_s > 0$ the skyrmion lattice 
transforms into the saturated state by the second-order transition of instability
type.
(b) The detail of the phase diagram (a) near the ordering temperature.
In figures (a) and (b) we indicate the transition lines for helicoids 
calculated in \cite{Mukamel85} (grey lines). 
(c) A fragment of a skyrmion lattice at low temperatures with dominating
rotational modes; (d) Near the ordering temperature a skyrmion lattice 
is characterized by strong spatial modulations of $m (\mathbf{r})$.
\label{fig:phasediagram}
}
\end{figure}

\begin{figure}
\includegraphics[width=0.9 \columnwidth]{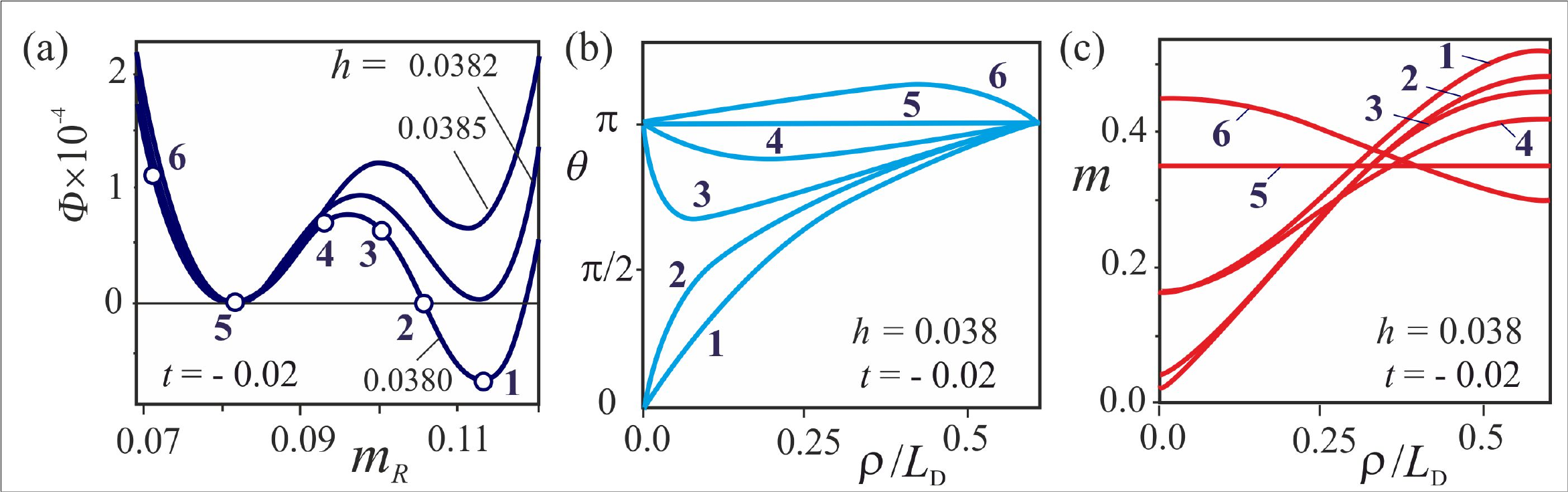}
\caption{ Energy density $\Phi (m_R)$ as a function of the boundary
value $m_R$ shows two minima corresponding to the skyrmion lattice 
(point \textbf{1}) and the saturated state (point \textbf{5}).
Magnetization profiles $\theta (\rho)$ (b) and $m (\rho)$ (c) drastically
change along the potential barrier $\Phi (m_R)$.
(color online).  
\label{fig:transition}
}
\end{figure}

Typical solutions for  hexagonal lattices with a fixed magnetization
modulus, i.e. well below the ordering temperature,
are collected in figure \ref{fig:unwinding} and for those in 
the precursor state  are presented in figures. \ref{fig:skyrmion2},
\ref{fig:phasediagram}, \ref{fig:transition}.
For the former a magnetic-field-driven evolution of the skyrmion lattice consists of two 
processes: an extension of the lattice period and a contraction of the skyrmion core 
\cite{JMMM94} (figure \ref{fig:unwinding}). At the critical field $H_{s} (T) = 0.801 H_D (T)$ 
the skyrmion lattice tranforms into the saturated state by an infinite extension 
of the lattice period   \cite{JMMM94}. 
Contrary to the common second-order transitions of Landau type, the lower symmetry
phase (skyrmion lattice) does not disappear completely during the transition but
 remains in the high symmetry phase (saturated state) in a form of localized states
(isolated skyrmions).  De Gennes  has designated these transitions as the
\textit{nucleation} type, and the common second-order transitions as the 
\textit{instability} type \cite{DeGennes75}.
The solutions for model (\ref{density2}) in the vicinity of
the ordering temperature ($t = -0.02$) (figure \ref{fig:skyrmion2}) demonstrate a dominant role
of longitudinal modulations. In this case the lattice periods practically do not change 
with increasing applied field and m($\mathbf{r}$) exhibits a profound spatial inhomogeneity
and a strong field dependence (figures \ref{fig:skyrmion2}).
 Despite the strong transformation of their internal structures 
the skyrmion lattices preserve \textit{axisymmetric} arrangement of the
magnetization near the centers of the lattice cells
(figure \ref{fig:skyrmion2} (a-d)). This remarkable property reflects 
the basic physical mechanism underlying the formation of skyrmion
lattices.
The local energetic advantage of Skyrmion lattices over helicoids is
due to a larger energy reduction in the ``double-twisted'' Skyrmion cell core 
compared to ``single-twisted'' helical states \cite{Nature06}.
This explains the unusual axial symmetry of the cell cores and their stability.
An increasing magnetic field gradually suppresses the antiparallel magnetization
in the cell core reducing the energetic advantage of the ``double-twist'' and 
increases the overall energy of the condensed skyrmion lattice.
At the critical field $h_f$, a first-order transition occurs into the saturated state. 
Above this field the skyrmion lattice still can exist as a metastable state 
until the magnetization modulus in the cell center becomes zero (see the magnetization profile 
for $h = 0.042$ in figure~\ref{fig:skyrmion2} (f)).

\textit{Phase diagrams}.
The calculated phase diagram (figure \ref{fig:phasediagram} (a)) shows
that the transition field $h_s(t)$ between the skyrmion lattice and saturated state has three
distinct sections separated by critical points $h_{s1}$ and $h_{s2}$.

(i) For $h_s (t) > h_{s1}$  the second-order transition of nucleation type occurs along 
the critical line $h_s(t)$. At lower temperatures where spatial variations of $m$ become
negligible this critical line transforms into transition line $h_s = 0.801H_D(T)/H_0$.

(ii) For $h_{s1} > h_s(t) > h_{s2}$  the skyrmion lattice transforms into the saturated
state by the first-order process. In this case skyrmion lattices can exist as metastable
state above the transition line. Figure \ref{fig:transition} (a) shows the potential
profiles $\Phi (m_R)$ in the vicinity of the first-order transition. 
Magnetization profiles $\theta (\rho/L_D)$, $m (\rho/L_D)$ 
in figures  \ref{fig:transition} (b) and (c) demonstrate the evolution
of solutions along the potential barrier.

iii) Finally, in the  vicinity of the ordering temperature
($h_{s2} > h_s(t) > 0$ ) the second-order process of instability type occurs.

Near the ordering temperature helicoids also consist of coupled rotational
and longitudinal modulations and exhibit a qualitatively similar evolution 
in the applied field \cite{Mukamel85,Yamashita87,Laliena16,Shinozaki17}. 
The phase transition lines between the helicoid and saturated state have been 
calculated in \cite{Mukamel85}.  In the phase diagram, they are indicated with
light grey colour (figure \ref{fig:phasediagram} (a), (b)). 

At lower temperatures ($ h_h(t) > h_{h1}$) 
the applied field gradually unwinds the helicoid into a set of isolated domain walls 
(\textit{kinks}) (the second-order transition of nucleation type).
In the temperature interval  $  h_{h1} > h_h(t) > h_{h2}$ the helicoid transforms
into the saturated phase by the first-order process, and near the ordering
temperature ($h_{h2} < h_h(t) < 0$) the second-order transition of instability type
occurs. Interestingly, the crossover of the inter-soliton coupling occurs on 
the same critical line as for skyrmions (\ref{criticalline1}). Detailed calculations
of helicoid states in the precursor state have been carried out by Shinozaki et al.
\cite{Shinozaki17}. 
Note that $h_h(t)$ line does not belong to the elements of the phase diagrams 
in figures \ref{fig:phasediagram} (a), (b), and are plotted to compare the helicoid ($h_h(t)$)
and skyrmion lattice ($h_s(t)$) transition fields.

Epilayers of cubic helimagnets with strong induced magnetic anisotropy of easy-plane type
\cite{Wilson14prb} and easy-plane noncentrosymmetric ferromagnets 
(e.g. a hexagonal helimagnet CrNb$_3$S$_6$ \cite{Togawa12PRL,Tsuruta16PRB,ClementsCrNbS}
provide an ideal setup for investigations of helicoid modulations in the regular and
precursor regions.
Chiral ferromagnets with easy-axis anisotropy and free-standing nanolayers of cubic helimagnets
are characterized by extended stability areas for helicoids and skyrmion lattices
\cite{Butenko10,Leonov16prl} and are convenient objects for investigations of
the precursor modulated states.

In the phase diagram (figure \ref{fig:phasediagram} (a)) we have plotted
line $h^{\star} (t)$ with \textit{turning point}  $p$ ($-1.0,\sqrt{2}/4$) 
and \textit{ending point} $u$ (-0.75, 0) described by the equation 
\cite{Leonov10condmat}
\begin{eqnarray}
h^{\star} =\sqrt{2(1  \pm  \sqrt{t+1})}(t+5/4 \pm \sqrt{t+1}), \quad t < -0.75
\label{criticalline1}
\end{eqnarray}
This line separates  the areas with different types of the skyrmion-skyrmion 
and kink-kink coupling \cite{Leonov10condmat,Mukamel85}.
 Isolated skyrmions with rotational modes are always repulsive \cite{JETPL95}. 
For skyrmions with  longitudinal and rotational modes above a certain critical temperature 
the inter-particle potential gains an oscillatory character and imposes an attractive 
coupling between skyrmions \cite{Leonov10condmat,Wilhelm11,Wilhelm12}.
Similar properties are exhibited by isolated chiral domain walls (kinks) \cite{Mukamel85}.
In the phase diagram (figure \ref{fig:phasediagram} (a)) the crossover fields
 of the phase transitions from the modulated (skyrmion lattice and helicoid) phases 
into the saturated states belong to the critical line $h^{\star} (t)$ (\ref{criticalline1}).
This fact demonstrates that the character of inter-particle coupling between isolated
skyrmions (kinks) determines the order of the phase transition from the skyrmion lattice (helicoid) 
into the saturated state. A skyrmion lattice composed of repulsive cores
transforms into the saturated states by the second-order transition of the nucleation type
(figure \ref{fig:unwinding}). A skyrmion lattice (helicoid) with attractive skyrmion cells transforms
into the saturated state at finite period by the first-order process (figure \ref{fig:skyrmion2}). 
In other words, the crossover of the inter-soliton coupling triggers the crossover of the
phase transition between the skyrmion lattice (helicoid) and saturated state. 
The first-order transition from the skyrmion lattice into the cone phase and
the formation of skyrmion clusters observed in cubic helimagnet Cu$_2$OSeO$_3$ \cite{Loudon17} 
are explained by the attractive inter-skyrmion potential in the cone phase \cite{Leonov16JPCM}.

The phase diagram of the solutions for skyrmion lattices and helicoids (figure \ref{fig:transition} (c))
surveys the soft modulus effects and crossover phenomena in the precursor states of cubic 
helimagnets. In bulk cubic helimagnets, however, the cone phase corresponds to the global 
minimum, and skyrmion lattices and helicoids are metastable \cite{Wilhelm12}. 
Skyrmion lattices are stabilized only in the A-phase, a small pocket in the vicinity of 
the ordering temperature. A coexistence of twisted and longitudinal modulations observed
in the A-phase \cite{Wilhelm12,Onose12prl} is a manifestation of soft modulus effects in
the skyrmion lattice phase. It was also established that in bulk MnSi,
the skyrmion lattice in the A-phase is saturated by the first-order process \cite{Pappas17prlmnsi}
in accordance with the phase diagram in figure \ref{fig:transition} (c).

Similar crossover phenomena occur in type-II superconductors. This group of superconductors
is usually characterized by the repulsive vortex-vortex potential and the nucleation type 
of the second order transition between Abrikosov vortex lattice phase and superconducting phase. 
However, for compounds with small values of the Ginzburg-Landau parameters the inter-vortex 
coupling has an oscillatory character \cite{Hubert71} which imposes the first-order transition 
between the vortex lattice and supercondting state \cite{Essmann71}.

 In conclusion, the equilibrium states of skyrmion lattices in the
precursor states of cubic helimagnets demonstrate the crossover from 
lower temperature twisting modulations (figure \ref{fig:unwinding}) 
to the precursor states with dominating longitudinal modes 
(figure \ref{fig:skyrmion2}).    
The calculated magnetic phase diagram  shows in the precursor region  
skyrmion lattices transform into the saturated phase by first-order processes 
(figure \ref{fig:phasediagram}, \ref{fig:transition}). These results 
are corroborated by the recent experimental findings \cite{XuAPL17FeGe,Pappas17prlmnsi}
and provide a theoretical basic for investigations of the precursor states in chiral 
ferromagnets.

  The authors grateful to J. Campo, K. Inoue, Y. Kato, J. Kishine, V. Laliena, Y. Masaki, 
	M. Mito, C. Pappas,  M. Shinozaki, and Y. Togawa for fruitful discussions and to
	J. Loudon for critical reading of the manuscript. AOL thanks Ulrike Nitzsche for
technical assistance.
	This work was funded by JSPS Core-to-Core Program, Advanced Research Networks (Japan)  and JSPS Grant-in-Aid for Research Ac-
tivity Start-up 17H06889.
%

\end{document}